\documentclass{article}
\usepackage{wds10,epsf}

\usepackage{graphicx}

\lefthead{SVOBODA ET AL.}
\righthead{COMPARISON OF RELATIVISTIC IRON LINE MODELS}

\newcommand{\laor}{{\sc laor}}
\newcommand{\kyrline}{{\sc{kyrline}}}
\newcommand{\powerlaw}{{\sc{powerlaw}}}
\newcommand{\diskbb}{{\sc{diskbb}}}
\newcommand{\phabs}{{\sc{phabs}}}

\newcommand{\pexrav}{{\sc{pexrav}}}
\newcommand{\pexriv}{{\sc{pexriv}}}
\newcommand{\smedge}{{\sc{smedge}}}
\newcommand{\refsch}{{\sc{refsch}}}

\begin{document}

\title{Comparison of relativistic iron line models}

\author{Ji\v{r}\'{i} Svoboda}

\affil{Astronomical Institute of the Academy of Sciences, Prague, Czech Republic\\ Charles University, Faculty  of  Mathematics  and  Physics,
     Prague, Czech Republic}

\author{Michal Dov\v{c}iak, Ren\'{e} W. Goosmann, Vladim\'{\i}r Karas}

\affil{Astronomical Institute of the Academy of Sciences, Prague, Czech Republic}

\begin{abstract}
The analysis of the broad iron line profile in the X-ray spectra of active galactic nuclei 
and black hole X-ray binaries allows us to constrain the spin parameter of the black hole.
We compare the constraints on the spin value for two X-ray sources, MCG-6-30-15 and GX 339-4, with a broad iron line 
using present relativistic line models in XSPEC --- {\laor} and {\kyrline}. 
The {\laor} model has the spin value set to the extremal value $a=0.9982$, while the {\kyrline} model enables
direct fitting of the spin parameter.
The spin value is constrained mainly by the lower boundary of the broad line,
which depends on the inner boundary of the disc emission where the gravitational redshift is maximal.
The position of the inner disc boundary is usually identified with the marginally stable orbit 
which is related to the spin value. In this way the {\laor} model can be used to estimate the spin value.
We investigate the consistency of the {\laor} and {\kyrline} models.
We find that the spin values evaluated by both models agree within the general uncertainties when applied on the current data. 
However, the results are apparently distinguishable for higher quality data, such as those simulated for the 
International X-ray Observatory (IXO) mission. We find that the {\laor} model tends to overestimate the spin value
and furthermore, it has insufficient resolution which affects the correct determination of
the high-energy edge of the broad line. \\


\end{abstract}

\begin{article}

\section{Introduction}

\begin{figure}[tbh!]
\begin{minipage}{0.62\textwidth}
\includegraphics[width=0.98\textwidth]{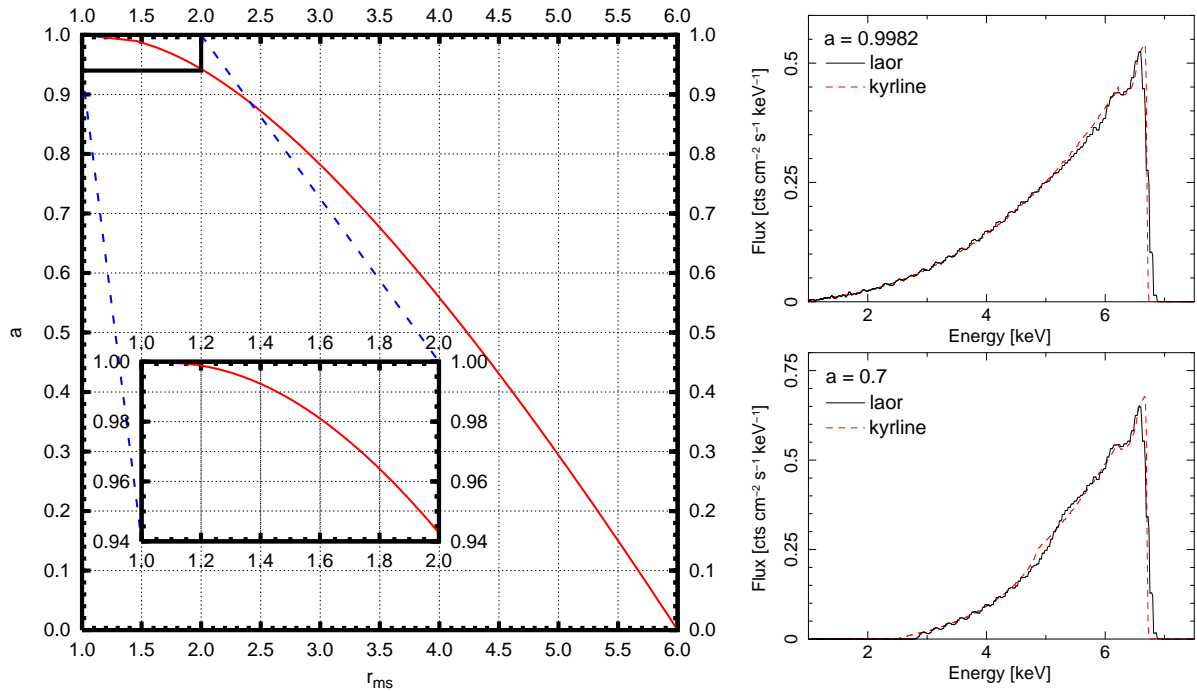}
\end{minipage}
\begin{minipage}{0.38\textwidth}
\includegraphics[angle=270,width=0.98\textwidth]{compare_laorky.eps} \\
\includegraphics[angle=270,width=0.98\textwidth]{compare_laorky_a07.eps} 
\end{minipage}
\caption{Left: Relation for the spin $a$ and marginally stable orbit $r_{\rm ms}$. Right: Comparison of the {\laor} (black, solid) and {\kyrline} (red, dashed) model for two values of the spin $a = 0.9982$ (top) and $a = 0.7$ (bottom). The other parameters of the line are $E=6.4$\,keV, $q_{1}=q_{2}=3$, $i=30^{\circ}$.}
\label{intro}
\end{figure}

The broad emission iron lines are well-known features found in about two dozens of spectra of active galactic nuclei and black hole binaries. 
They are supposed to originate close to the black hole by the reflection of the primary radiation
on the accretion disc.
The spin of the black hole plays an important role in the forming of the line shape. Especially,
it determines the position of the marginally stable orbit which is supposed to confine the inner edge of the accretion disc
(see Figure~\ref{intro}). The innermost stable orbit occurs closer to a black hole with a higher spin value. However, the spin affects also the overall shape of the line. 

Over almost two decades the most widely used model of the relativistic disc spectral line has been the one by \markcite{{\it Laor} [1991]},
which includes the effects of a maximally rotating Kerr black hole. In other words, the {\laor} model sets the dimensionless angular momentum $a$ to the canonical value of $a=0.9982$ -- so that it cannot be subject of the data fitting procedure. 
\markcite{{\it Dov\v ciak et al.} [2004]} have relaxed this limitation and allowed $a$ to be fitted in the suite of {\sc ky} models.
Other numerical codes have been developed independently by several groups (\markcite{{\it Beckwith and Done} [2004]}, \markcite{{\it \v{C}ade\v{z} and Calvani} [2005]}, \markcite{{\it Brenneman and Reynolds} [2006]}) and equipped with similar functionality.

However, the {\laor} model can still be used for evaluation of the spin if one identifies the inner edge of the disc with the marginally stable orbit. In this case the spin is actually estimated from the lower boundary of the broad line. 
The comparison of the {\laor} and {\kyrline} model is shown in the right panel of Figure~\ref{intro}.
The other parameters of the relativistic line models are inclination angle $i$, rest energy of the line $E$, inner radius of the disc $R_{\rm in}$, outer radius of the disc $R_{\rm out}$, emissivity parameters $q_{1}$, $q_{2}$ with the break radius $r_{b}$. The emissivity of the line is given by $I \approx r^{-q_{1}}$ for $r<r_{b}$ and $I \approx r^{-q_{2}}$ for $ r>r_{b}$. The angular dependence of the emissivity is characterized by limb darkening profile $I(\mu_{\rm e})\propto 1 + 2.06\mu_{\rm e}$ in the {\laor} model.
The {\kyrline} model enables to switch between different emission laws. We used further two extreme cases, the {\kyrline} with the same limb-darkening law as in the {\laor} model and {\kyrline}* with the limb-brightening law $I(\mu_{\rm e}) \propto \ln (1+\frac{1}{\cos i})$.

The aim of this paper is to compare the two models applied to the current data provided by the XMM-Newton satellite, and to the artificial data generated for the on-coming X-ray mission.
For this purpose we have chosen two sources, MCG-6-30-15 and GX 339-4, which exhibit an extremely skewed iron line
according to recently published papers (\markcite{{\it Vaughan and Fabian} [2004], {\it Miller et al.} [2004]}).


\section{Observations and data reduction}

We used the SAS software version 7.1.2 (http://xmm.esac.esa.int/sas) to reduce the XMM-Newton data of the sources. Further, we used standard tools for preparing and fitting the data available at http://heasarc.gsfc.nasa.gov (FTOOLS, XSPEC) 

The galaxy MCG-6-30-15 is a nearby Seyfert 1 galaxy ($z = 0.008$). The skewed iron line has been revealed in the X-ray spectra 
by all recent satellites. The XMM-Newton observed MCG-6-30-15 for a long 350\,ks exposure time during summer 2001 (revolutions 301, 302, 303). The spectral results are described in \markcite{{\it Fabian et al.} [2002]}. We joined the three spectra into one using the ftool MATHPHA.

The black hole binary GX 339-4 exhibited a strong broadened line in the 76\,ks observation in 2002 (\markcite{{\it Miller et al.} [2004]})
when the source was in the very high state (for a description of the different states see \markcite{{\it Remillard and McClintock} [2006]}). The observation was made in the \textit{burst mode} due to a very high source flux. The 97$\%$ of photons are lost during the reading cycle in this mode, which results into 2.25\,ks total exposure time.\footnote{The broad iron line was also revealed in the analysis of the two 138\,ks observations in spring 2004 by \markcite{{\it Miller et al.} [2006]}, when the source was in the low-hard state. The EPIC pn camera was operating in the \textit{timing mode}, MOS cameras in the \textit{full-frame mode}. Using \textit{epatplot} tool we found that it is not possible to avoid the pile-up by excluding the central part of the image of the source as described in the forementioned paper. Therefore we use only the very high state observation from 2002 in our analysis.} 

We rebinned all the data channels in order to oversample the instrumental energy resolution maximally by a factor of 3 and to have at least 20 counts per bin. The first condition is much stronger with respect to the total flux of the sources -- $4\times10^{-11}$\,erg\,cm$^{-2}$\,s$^{-1}$ in 2--10\,keV ($1.1\times10^{6}$\,cts) for MCG-6-30-15 and $9\times 10^{-9}$\,erg\,cm$^{-2}$\,s$^{-1}$ in 2--10\,keV ($1.0\times10^{7}$\,cts) for GX 339-4.



\section{Iron line study of the MCG-6-30-15 spectrum}

\begin{figure}[tbh!]
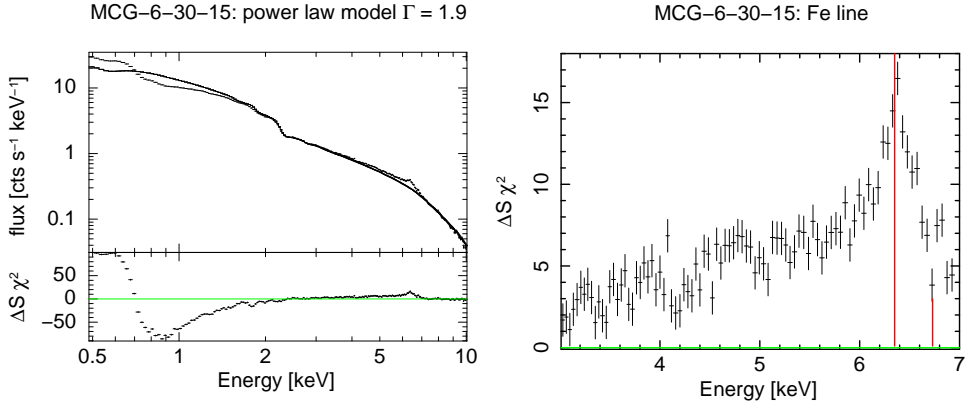

\begin{center}
\begin{tabular}{ccc}
\includegraphics[angle=270,width=0.4\textwidth]{mcg_powerlaw.eps}
\includegraphics[angle=270,width=0.4\textwidth]{mcg_ironlineregion.eps}
\end{tabular}
\caption{The X-ray spectrum of MCG-6-30-15 observed by XMM-Newton. Left: The overall view on the spectrum as a power law with $\Gamma=1.9$ absorbed by a neutral hydrogen along the line of sight with column density $n_{\rm H} = 0.41\times 10^{21}$\,cm$^{-2}$. The residuals from 
the model are plotted in the bottom panel clearly revealing features of a local absorption and soft excess at the soft X-ray band,
and a feature at around $6$\,keV which can be explained by the presence of a broad iron line. Right: More detailed view of the iron line band.}
\label{mcg}
\end{center}
\end{figure}

We used the same continuum model for the MCG-6-30-15 spectrum as presented in \markcite{{\it Fabian et al.} [2002]}: the simple power law component absorbed by neutral hydrogen with the column density $n_{\rm H} = 0.41\times 10^{21}$\,cm$^{-2}$.
The overall spectrum with a detailed view of the iron line energy band is shown in Figure~\ref{mcg}.  The employed model is sufficient to fit the data above $\approx 2.5$\,keV, which is satisfactory for our goal of the comparison of the relativistic line models.\footnote{Another components are needed to be added into the model in order to fully understand the spectrum. Several works have been done in this way, in the most recent one by \markcite{{\it Miller~L. et al.} [2008]} the spectrum is characterized by an absorption in four different zones, which affects also the higher energy band where no broad line is needed any more.}
The value of the photon index is $\Gamma =1.90(1)$. The residuals are formed by a complex of a broad iron line and two narrow iron lines --
one emission line at $E=6.4$\,keV likely originating in a distant matter (torus) and one absorption line at $E = 6.77$\,keV which
can be explained by a blueshifted absorption originating in an outflow.
The rest energy of the broad line is $E=6.7$\,keV, which corresponds to the helium-like ionized iron atoms.\footnote{The spectral complexity in the line band allows an alternative explanation -- the model with two narrow emission lines at energies $E=6.4$\,keV and $E=6.97$\,keV. This alternative model leads to the presence of the broad line component at $E=6.4$\,keV.}

A good fit of the broad line was found with a broken power law line emissivity with a steeper dependence on the radius in the innermost region, which suggests a centrally localized corona. The goodness of the fit is constrained by the least squared method.
The fit results in 2.5--9.5\,keV  are presented in Table~1. 
The $\chi^{2}$ values give comparable results for all employed models. The $\chi^{2}_{\rm red}=\chi^{2}/\nu \approx 1.2$, where $\nu$ 
is the number of degrees of freedom which is related to the total number of energy bins and model parameters.
The six independent parameters of the {\laor} and {\kyrline} models make the global minimum of $\chi ^{2}$ rather wide with several local minima. Each model has a different tendency to converge to a different minimum.
Hence, we did not compare only best fits of both models, but also the evaluated spin values by the {\kyrline} and {\laor} models
when the other model parameters correspond to each other.
The equivalent width of the line is $EW \approx 750$\,eV. The errors in brackets presented in the table correspond to 90$\%$ confidence
and are evaluated when the other parameters of the model are fixed. The realistic errors are higher because the model parameters 
further depend on the other parameters of the line and continuum models.

\begin{figure}[tbh!]
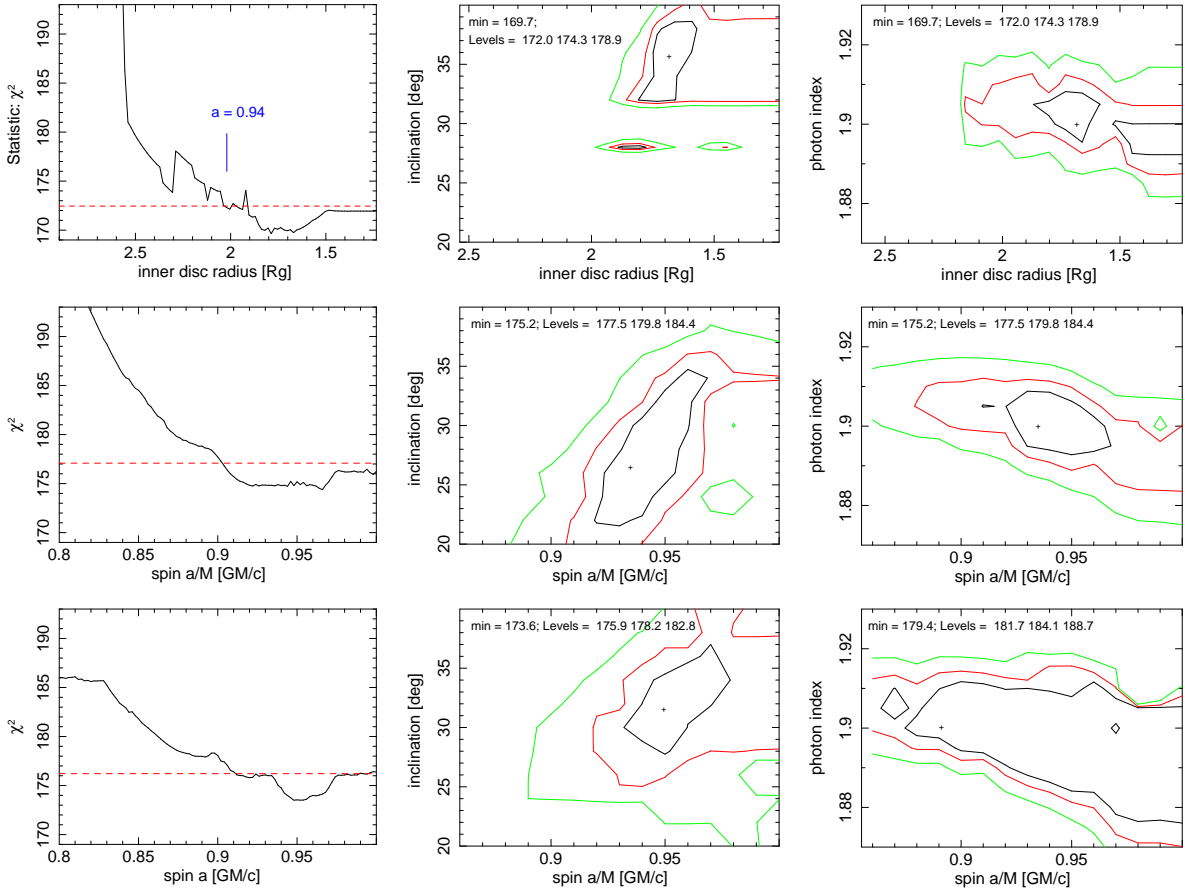

\begin{center}
\begin{tabular}{ccc}
  \includegraphics[angle=270,width=0.31\textwidth]{laor_stepa2.eps} &
  \includegraphics[angle=270,width=0.31\textwidth]{laor_contirin2.eps} &
  \includegraphics[angle=270,width=0.31\textwidth]{laor_contpa2.eps}\\
  \includegraphics[angle=270,width=0.31\textwidth]{kyr_stepa2.eps} &
  \includegraphics[angle=270,width=0.31\textwidth]{kyr_contia2.eps} &
  \includegraphics[angle=270,width=0.31\textwidth]{kyr_contpa2.eps}\\
  \includegraphics[angle=270,width=0.31\textwidth]{kylb_stepa.eps} &
  \includegraphics[angle=270,width=0.31\textwidth]{kylb_contia1.eps} &
  \includegraphics[angle=270,width=0.31\textwidth]{kylb_contpa2.eps}\\
\end{tabular}
\caption{The contour graphs show the dependence of the value of the $\chi ^{2}$ , the inclination angle and the power law index on the value of the spin ({\kyrline}) or the inner disc radius ({\laor}) for the MCG-6-30-15 spectrum in 2.5--9.5\,keV. The black, red and green contours correspond to $1\sigma$, $2\sigma$ and $3\sigma$, respectively. Top: The results of the {\laor} model.
Middle: The results of the {\kyrline} model with \textit{limb darkening}. Bottom: The results of the {\kyrline} model with \textit{limb brightening}.}
\label{mcg_contours}
\end{center}
\end{figure}

To catch up these relations we produce various contour graphs focusing on the determination of the spin value, taking into account the other parameters of the used model. The relations of the $\chi ^{2}$ values on value of the spin ({\kyrline}) or the inner disc radius ({\laor}) 
are shown in the left column of Figure~\ref{mcg_contours}. 
The x-axis is oppositely directed in the case of the inner disc radius as x-variable for an easier comparison with the {\kyrline} results.
The contour graphs for the spin and the inclination angle are shown in the middle column of Figure~\ref{mcg_contours}. The underlying model was fixed in both cases. 
The plots in the right column of Figure~\ref{mcg_contours} show the contours for the spin and the power law index. 
Taking all of these into account, we obtain for the spin value:\\
\[
a_{KY} = 0.94^{+0.06}_{-0.10}
\hspace{0.3cm} \rm{and} \hspace{0.3cm}
a_{laor} = 0.96^{+0.04}_{-0.08}.
\]

\vspace{0.2cm}

\begin{center}
\begin{tabular}{c|c|c|c|c}
 	\multicolumn{5}{c}{\bf Table 1. Results for MCG-6-30-15 in 2.5--9.5\,keV}\\
	\hline \hline
	parameter   &	{\kyrline}&	{\kyrline}*	& {\laor}\,$_{\rm best}$	 & {\laor}\,$_{\rm loc.min.}$\\
	\hline
	\rule{0cm}{0.35cm}
	$a/M$		&	$\textbf{0.94}^{+0.02}_{-0.03}$&	$\textbf{0.95}^{+0.02}_{-0.01}$	&	$\textbf{0.98}^{+0.02}_{-0.01}$	&	$\textbf{0.96}^{+0.02}_{-0.01}$	\\
	$i$\,[deg]	&	$26.7(7)$		&	$31.5(7)$		&	$35.7(5)$		&	$26.8(5)$	\\
	$E$\,[keV]	&	$6.67(1)$		&	$6.60(1)$		&	$6.48(1)$		&	$6.66(1)$		\\
	$q_{1}$		&	$4.9(1)$		&	$3.7(1)$		&	$4.8(1)$		&	$4.7(1)$		\\
	$q_{2}$		&	$2.84(4)$		&	$2.11(4)$		&	$2.50(3)$		&	$2.87(3)$		\\
	$r_{b}$ 	&	$5.5(2)$		& 	$18.3(5)$		& 	$6.6(2)$		& 	$5.1(2)$		\\
	\hline
	$\chi^{2}/v$	&	$175/148$	&	$174/148$ 	&	$170/148$ 	&	$174/148$ 	\\
	\hline
	$EW$\,[eV]	&	$761$		&	$757$	  	&	$764$	  	&	$754$	  	\\
\end{tabular}
\end{center}


\section{Iron line study of the GX 339-4 spectrum}

\begin{figure}[tbh!]
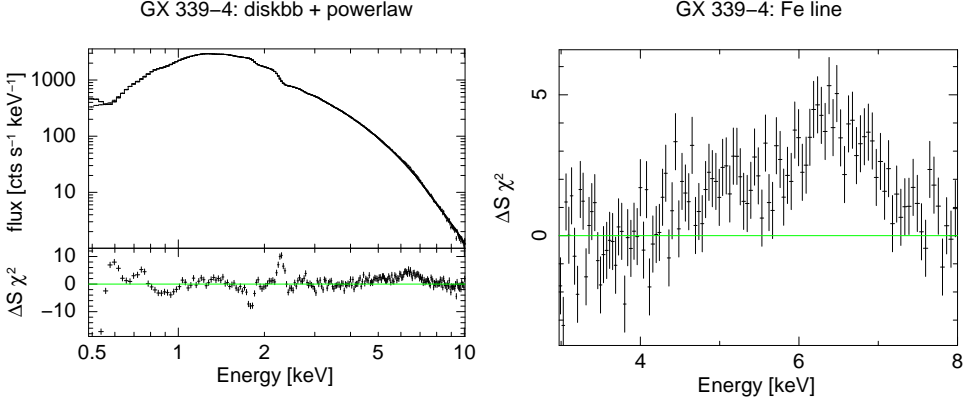

\begin{center}
\begin{tabular}{ccc}
\includegraphics[angle=270,width=0.4\textwidth]{gx_cont.eps}
\includegraphics[angle=270,width=0.4\textwidth]{gx_ironlineregion.eps}
\end{tabular}
\caption{The X-ray spectrum of GX 339-4. Left: The overall view on the spectrum as a power law with $\Gamma=3$ and thermal multi
temperature black body with $0.87$\,keV absorbed by a neutral hydrogen with column density $n_{\rm H} = 0.6\times 10^{22}$\,cm$^{-2}$. 
The residuals from the model are plotted in the bottom panel revealing wiggles at the soft X-ray band which are likely due
to instrumental response. 
The broad excess at around $6$\,keV can be explained by the presence of a broad iron line. 
Right: More detailed view of the iron line band.}
\label{gx}
\end{center}
\end{figure}

The continuum of the X-ray spectrum of the black hole binaries is characterized by a power law and a multi-colour disc black-body component 
({\powerlaw}\,+\,{\diskbb} in the XSPEC notation). 
The power law index suggested from the simultaneous RXTE measurements is $\Gamma = 2.5$ (\markcite{{\it Miller et al.} [2004]}). However, we get an unacceptable fit with $\chi^{2}_{\rm reduced} \ge 6.7$ using the same model as in the analysis by \markcite{{\it Miller et al.} [2004]} or the re-analysis by \markcite{{\it Reis et al.} [2008]}. The difference of the results is likely due to a different grouping of the instrumental energy channels applied to the data. While we did not allow to oversample the instrumental energy resolution more than by factor of 3, in the previous works only the condition to have at least 20 counts per bin was used. This condition is very weak with respect to the total number of counts $N_{\rm counts} \approx 1.0\times 10^{7}$ and the total number of energy channels $N_{\rm chan} = 1.5\times 10^{3}$ in 2--10\,keV and as a result, it practically does not force the data to be grouped. This leads to an excessive oversampling of the energy resolution, to large error bars in the flux and finally to an artificial decrease of the $\chi^{2}_{\rm red}$ value.
In the energy range 0.8--9\,keV we get $\chi^{2}/\nu = 2835/1640$ for the grouping with the only condition 20 counts per bin,
and $\chi^{2}/\nu = 1368/202$ for the grouping taking the energy resolution into account. 
The $\chi^{2}_{\rm red}$ value increased from $\chi^{2}_{\rm red} \doteq 1.73$ to $\chi^{2}_{\rm red} \doteq 6.77$.

The absorption and emission features around $E\approx 2$\,keV have a significant effect on the goodness of the fit.
These features can be linked with the Si and Au edges suggesting to be instrumental imprints. Hence, we added two gaussian lines into the model to improve the fit ($\chi^{2}/\nu \rightarrow 667/202$). As a next step, we allowed the continuum parameters to float (and also removed the {\smedge} model used in the previous works). The new model has $\chi^{2}/\nu = 350/202$ in 0.8--9\,keV and its parameter values are $n_{\rm H} = 0.6\times 10^{22}$\,cm$^{-2}$, $\Gamma = 3.08(5)$ and $kT_{\rm in} = 0.87(1)$\,keV. We tried to add a reflection component into the model, as {\pexrav} ({\pexriv}) or {\refsch} model, but without any improvements of the fit. 
%
%
The spectrum of GX 339-4 is shown in Figure~\ref{gx} with a detailed view of the iron line band in the right panel. 
A broadened iron line feature is still present. However, due to different adopted value for the photon index of the power law
the line is much weaker than the one presented in \markcite{{\it Miller et al.} [2004]}.

\begin{figure}[tbh!]
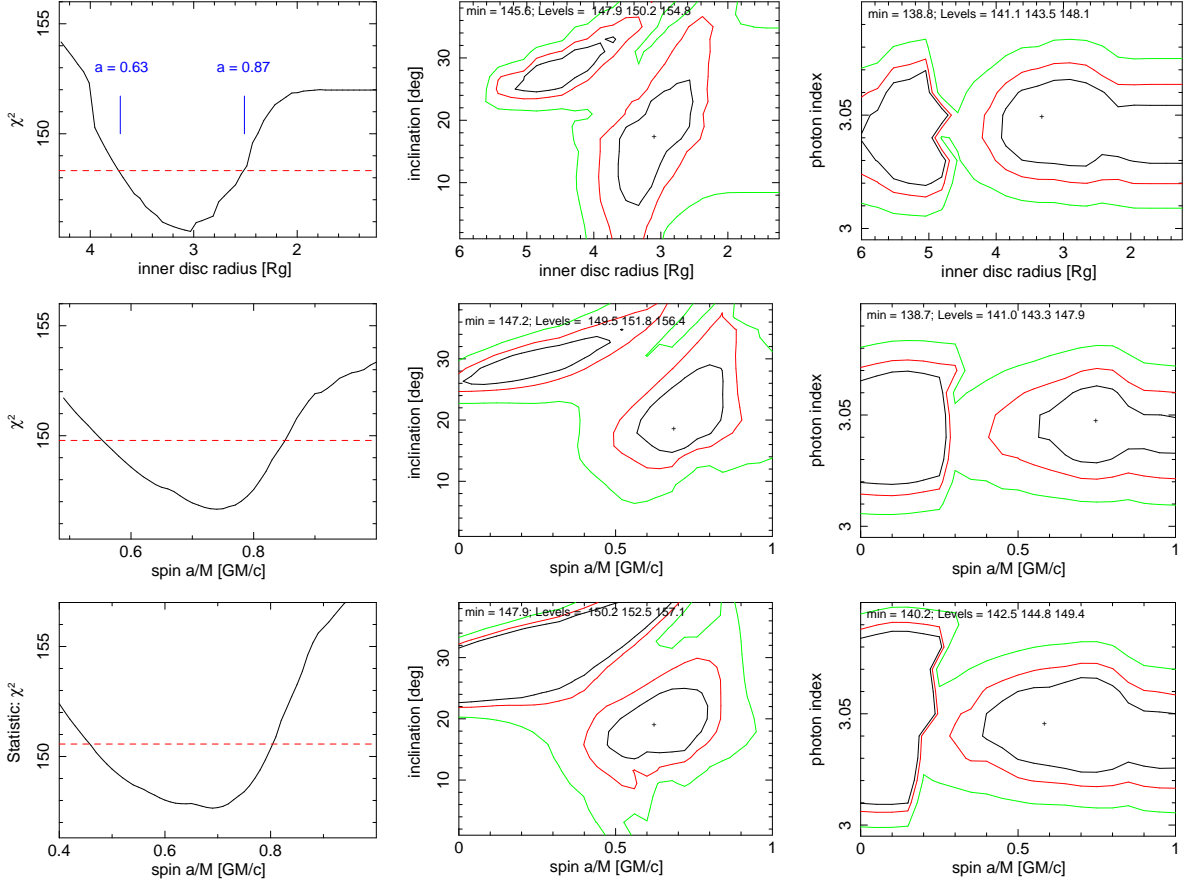

\begin{center}
\begin{tabular}{ccc}
 \includegraphics[angle=270,width=0.31\textwidth]{gxlaor_stepa2.eps} &
 \includegraphics[angle=270,width=0.31\textwidth]{gxlaor_contirin.eps} &
 \includegraphics[angle=270,width=0.31\textwidth]{gxlaor_contpa.eps}\\
 \includegraphics[angle=270,width=0.31\textwidth]{gxkyr_stepa.eps} &
 \includegraphics[angle=270,width=0.31\textwidth]{gxkyr_contia.eps} &
 \includegraphics[angle=270,width=0.31\textwidth]{gxkyr_contpa.eps}\\
 \includegraphics[angle=270,width=0.31\textwidth]{gxkylb_stepa.eps} &
 \includegraphics[angle=270,width=0.31\textwidth]{gxkylb_contia.eps} &
 \includegraphics[angle=270,width=0.31\textwidth]{gxkylb_contpa.eps}\\
\end{tabular}
\caption{The contour graphs show the dependence of the value of the $\chi ^{2}$, the inclination angle and the power law index on the value of the spin ({\kyrline}) or the inner disc radius ({\laor}) for the GX 339-4 spectrum in 3--9\,keV. The black, red and green contours correspond to $1\sigma$, $2\sigma$ and $3\sigma$, respectively. Up: The results of the {\laor} model.
Middle: The results of the {\kyrline} model with \textit{limb darkening}. Down: The results of the {\kyrline} model with \textit{limb brightening}.}
\label{gx_contours}
\end{center}
\end{figure}

The fitting results of the line models in 3--9\,keV are summarized in Table 2 and Figure~\ref{gx_contours}.
There are two minima found during the fitting procedure. We preferred the one which better corresponds to the results
obtained by the independent radio and infrared measurements which constrained the inclination angle to be $i < 26^{\circ}$
(\markcite{{\it Gallo et al.} [2004]}).
The dependence of the goodness of the fit on the spin value is shown in the left column of Figure~\ref{gx_contours}. 
The contour graphs for the spin and the inclination angle are depicted in the middle column, and for the spin 
and the power law photon index in the right column of Figure~\ref{gx_contours}. 
The derived spin value is then:
\[
a_{KY} = 0.69^{+0.16}_{-0.13}
\hspace{0.3cm} \rm{and} \hspace{0.3cm}
a_{laor} = 0.77^{+0.10}_{-0.14}.
\]


\begin{minipage}{0.5\textwidth}
\begin{table}[tbh!]
\begin{tabular}{c|c|c|c}
	\hline \hline
	parameter   &	{\kyrline}&	{\kyrline}*	& {\laor}	 \\
	\hline
	\rule{0cm}{0.35cm}
	$a/M$		&	$\textbf{0.69}^{+0.13}_{-0.12}$&	$\textbf{0.62}^{+0.14}_{-0.14}$	&	$\textbf{0.77}^{+0.08}_{-0.12}$\\
	$i$\,[deg]	&	$19(3)$		&	$19(4)$		&	$17(4)$		\\
	$E$\,[keV]	&	$6.97(1)$		&	$6.97(1)$		&	$6.97(1)$		\\
	$q$		&	$3.45(8)$		&	$3.35(8)$		&	$3.3(1)$		\\
	\hline
	$\chi^{2}/v$	&	$147/125$	&	$148/125$ 	&	$148/125$	\\
	\hline
	$EW$\,[eV]	&	$175$		&	$164$	  	&	$164$	  	\\
\end{tabular}
\caption{Results for GX 339-4 in 3--9\,keV}
\end{table}
\end{minipage}
\begin{minipage}{0.5\textwidth}
\begin{table}[tbh!]
\begin{tabular}{c|c|c}
	\hline \hline
			&	\small{MCG-6-30-15}		&	\small{GX 339-4}			\\
	\hline
	\rule{0cm}{0.35cm}
	net cts/s	&	$3.59$			&	$592.1$			\\
	model cts/s	&	$3.59$			&	$592.5$	\\
	line cts/s	&	$0.20$			&	$5.1$			\\
        line cts	&	$4.37\times10^{4}$		&	$1.15\times10^{4}$			\\
\multicolumn{3}{c}{}\\
\multicolumn{3}{c}{}\\
\end{tabular}
\caption{Count rates of the observations}
\end{table}
\end{minipage}

\begin{figure}[tbh!]
\begin{center}
\begin{tabular}{ccc}
\includegraphics[width=0.41\textwidth]{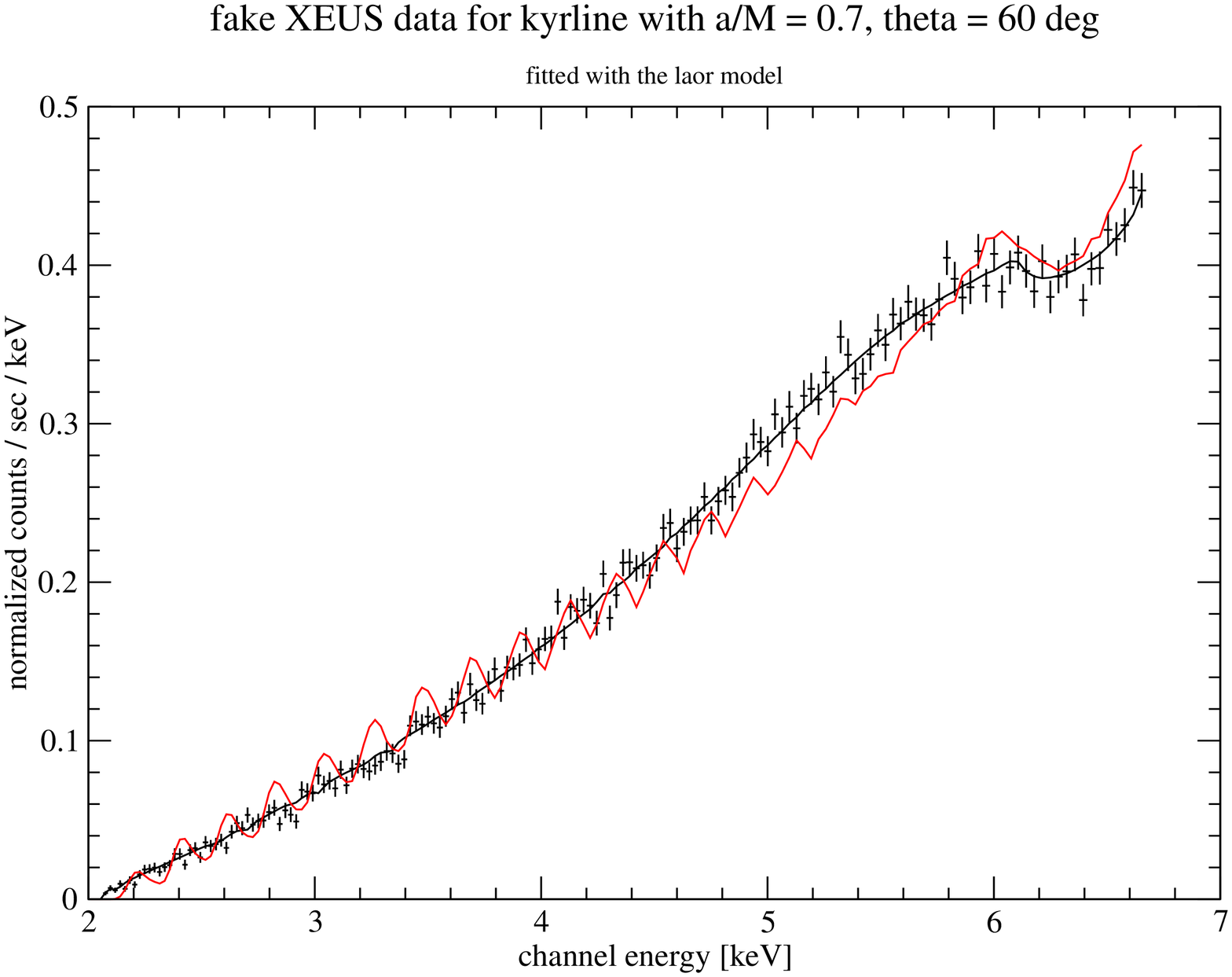} &
\includegraphics[width=0.41\textwidth]{a07-mcg-6-kyrline-alpha3-theta80-1-9keV-res30eV-laor-fit.eps} \\[-0.73cm]
\includegraphics[width=0.41\textwidth]{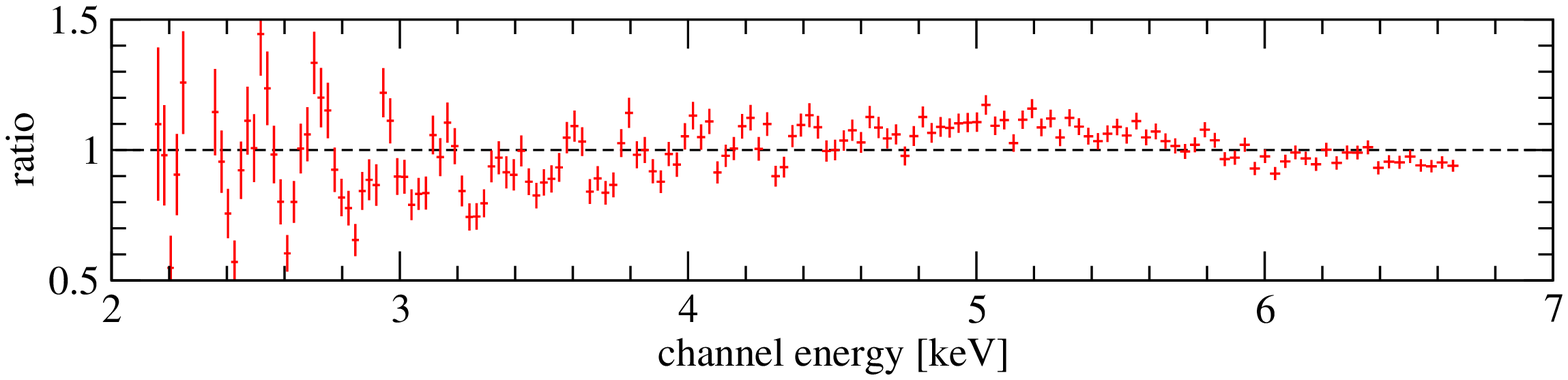} &
\includegraphics[width=0.41\textwidth]{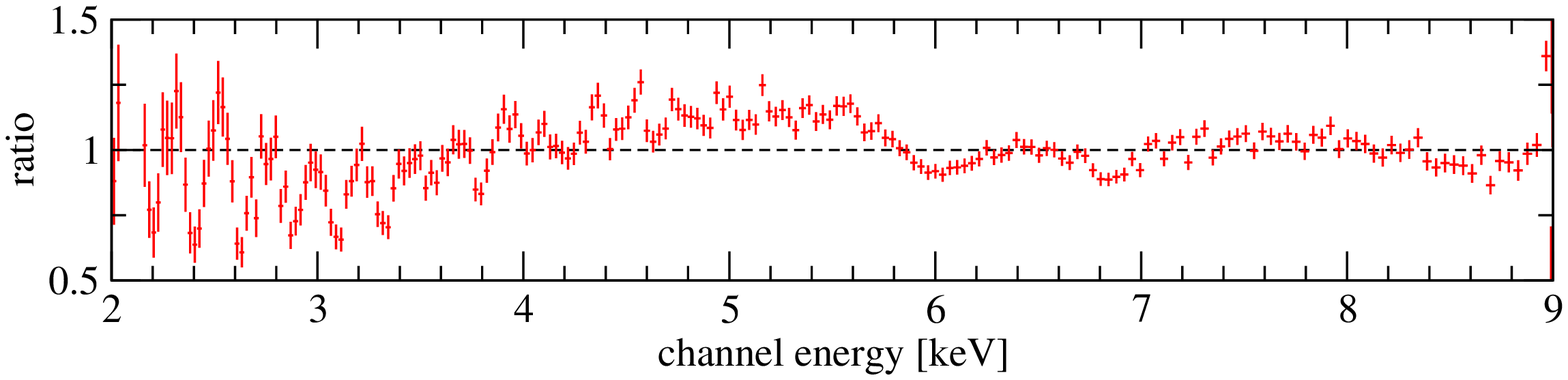}\\[5mm]
\includegraphics[width=0.41\textwidth]{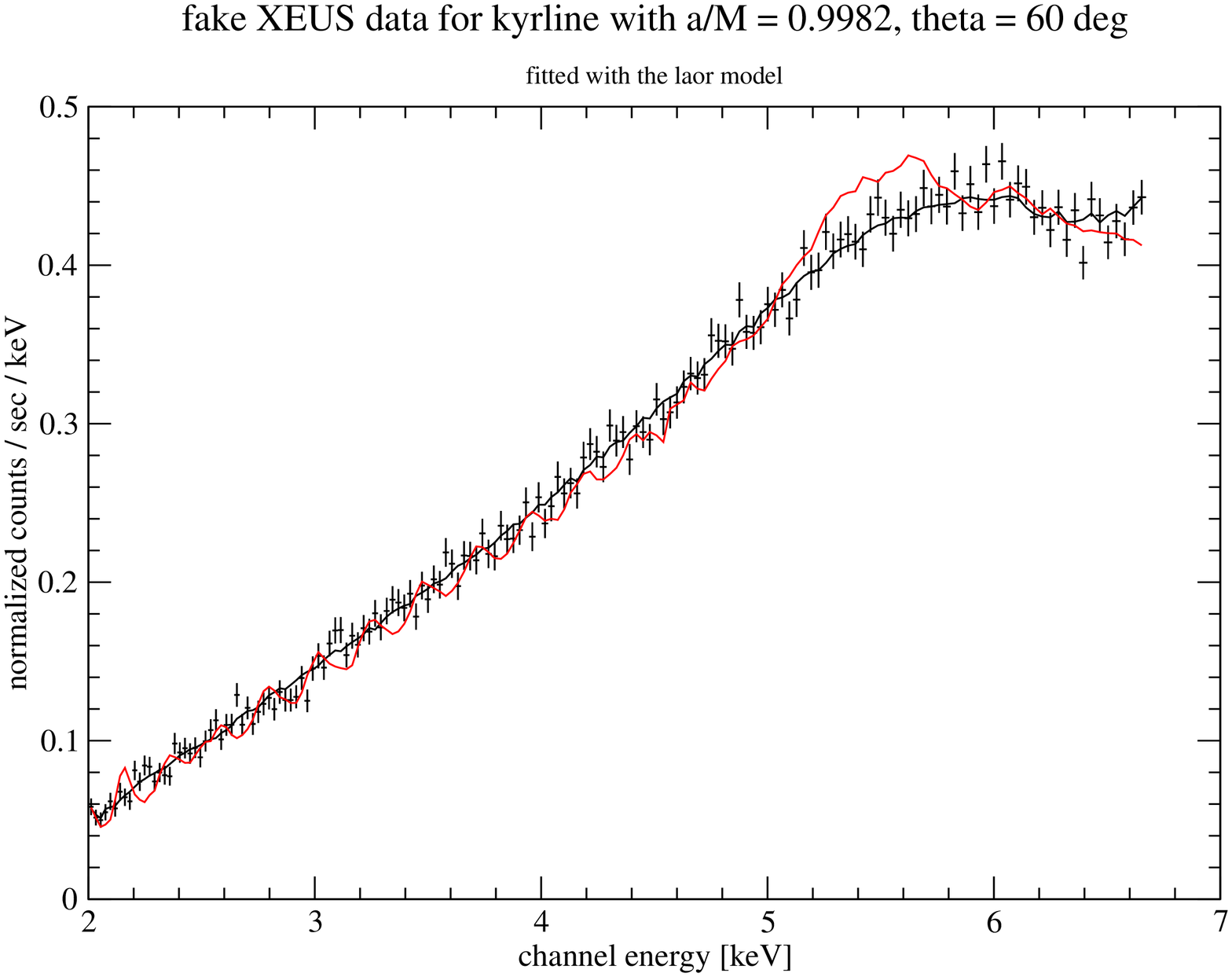} &
\includegraphics[width=0.41\textwidth]{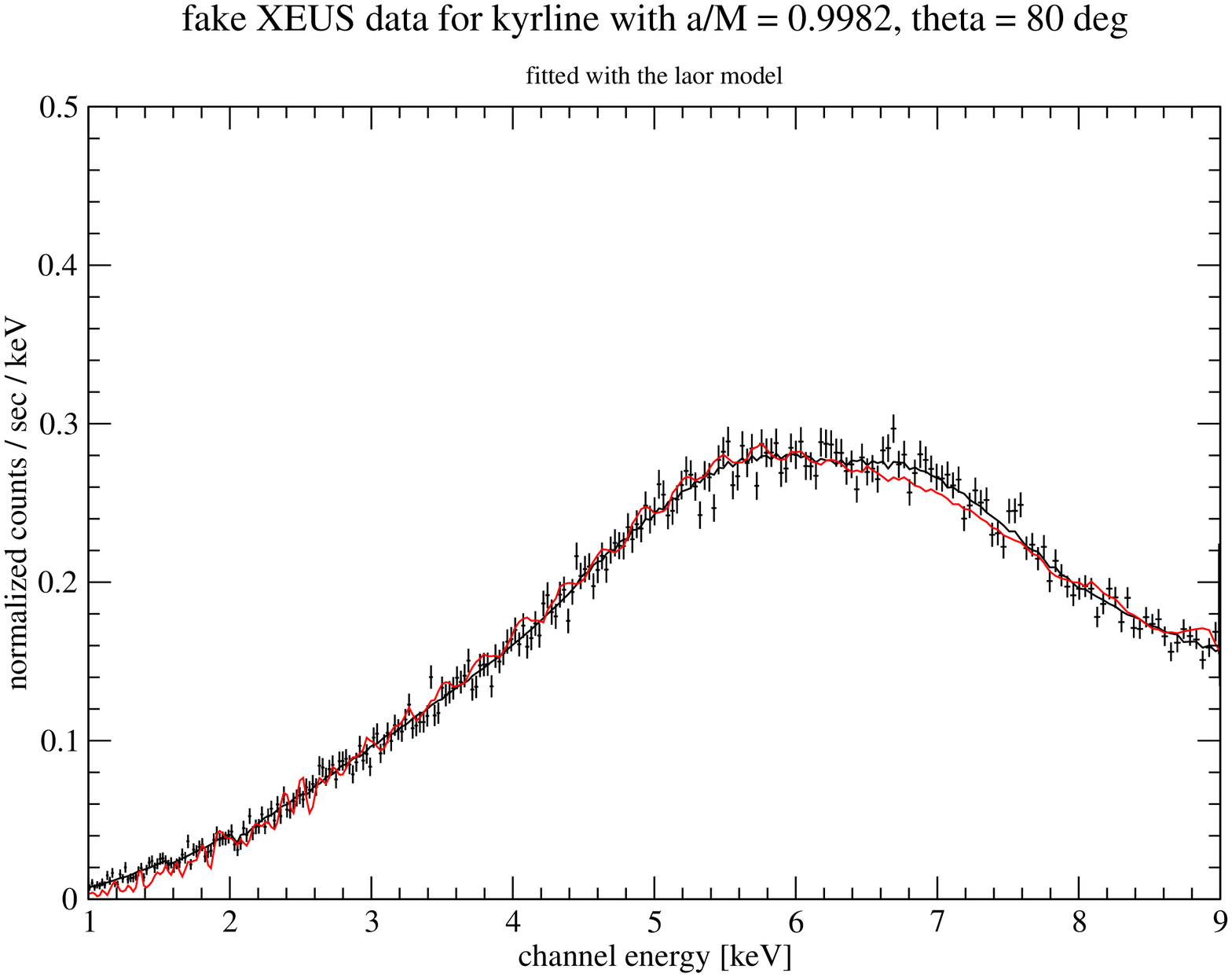} \\[-0.73cm]
\includegraphics[width=0.41\textwidth]{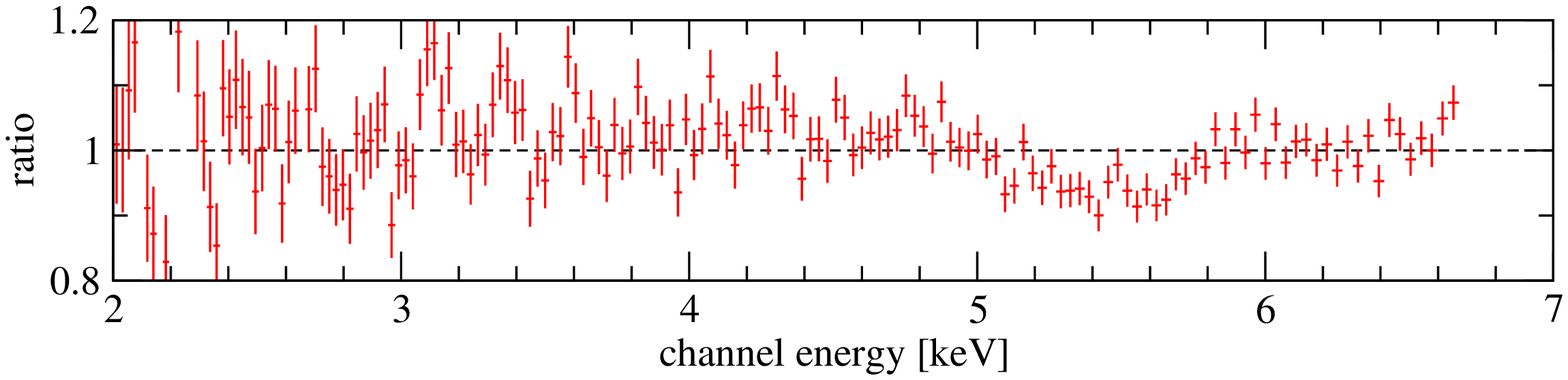} &
\includegraphics[width=0.41\textwidth]{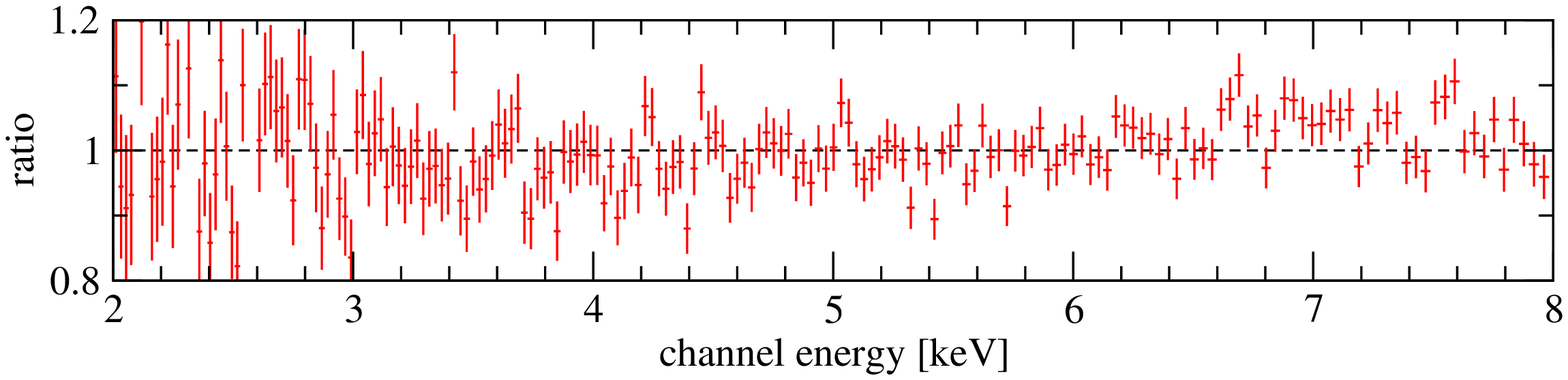}\\[5mm]
\end{tabular}
\caption{Simulated data for two values of the spin and inclination using the preliminary response matrix of the XEUS mission. We show the artificial data with the expected errors (black crosses) which were simulated by the {\kyrline} model (black curve), and fitted by the {\laor} model (red curve).
}
\label{xeus_systematic}
\end{center}
\end{figure}

\section{Fitting of the simulated data of future X-ray missions}
\label{future}

In this section we intend to apply the {\laor} and {\kyrline} models on the data 
with significantly higher quality supposed to be achieved by on-coming X-ray missions. 
The presently planned International X-ray Observatory (IXO) arised from the merging
of the former XEUS and Constellation-X missions. Because the details of the IXO mission have not been
fixed yet, we used a preliminary response matrix of the former XEUS mission (\markcite{{\it Arnaud et al.} [2008]}).
We generated the data for a {\kyrline} model with a rest energy of the line $E=6.4\,$keV
and radial disc luminosity that follows a power law with the index $q=3$. 
We rebinned the data in order to have a resolution of $30\,$eV per
bin. We then fit the data in the 1--9\,keV energy range with the {\laor} model using the same initial
values of the fitting parameters as for the data simulation. 
Due to insufficient resolution of the {\laor} model, a significant problem appears 
at the high-energy edge of the broad line. This occurs because the next generation instruments
achieve much higher sensitivity in the corresponding energy range.
Therefore, we excluded the higher-energy drop of the lines from the
analysis in order to reveal the differences in the overall shape of the line. 
We examined the artificial data for a grid of values of the angular momentum and the inclination angle.
The results for $a/M = 0.7$, $0.9982$ and $i = 60^{\circ}$, $80^{\circ}$ are shown in Figure~\ref{xeus_systematic}.
It is clearly seen from the figure that the effect of the spin on the shape of the line is sufficiently
resolved by the higher quality data. 

Further, we produced simulated data for the Seyfert galaxy MCG-6-30-15 and the
black hole binary GX 339-4 using rather simplified models which were suitable to fit the current XMM-Newton data. 
For MCG-6-30-15 we used a power law model plus a {\kyrline} model
for the broad iron line, absorbed by neutral hydrogen: {\phabs} * ({\powerlaw} + {\kyrline}). The
parameters of the continuum are the column density $n_{\rm H} = 0.4\times
10^{21}$\,cm$^{-2}$, the photon index $\Gamma =1.9$ of the power law, and its
normalization $K_{\Gamma} = 5\times 10^{-3}$. 
The values of the line parameters are summarized in the \textit{KY value} column of Table~4.
The exposure time was chosen as $220$\,ks and the flux of the source as $1.5\times10^{-11}$erg\,cm$^{-2}$\,s$^{-1}$ in the 2--10\,keV 
energy range (i.e. $1.4\times10^{7}$\,cts).

For GX 339-4 we used {\phabs} * ({\diskbb} + {\powerlaw} + {\kyrline}) with $n_{\rm H} =
0.6\times 10^{22}$\,cm$^{-2}$, $kT_{\rm in} = 0.87$\,keV ($K_{\rm kT} = 1.4\times 10^{3}$), and $\Gamma = 3$ ($K_{\Gamma} = 5.6$). 
The exposure time was chosen as $75$\,ks, the flux of the source as $9.3\times10^{-9}$erg\,cm$^{-2}$\,s$^{-1}$ in the 2--10\,keV 
energy range (i.e. $1.5\times10^{9}$\,cts).\footnote{The number of counts is two orders of magnitude higher than for the observation of the XMM-Newton satellite. The reason is due to the loss of 97\% of the photons during the \textit{burst mode} of XMM-Newton observation which eliminates the pile-up problem. The next generation X-ray missions are supposed to have a calorimeter instead of the CCD camera on-board which will get rid of such problems.}

\begin{figure}[tbh!]
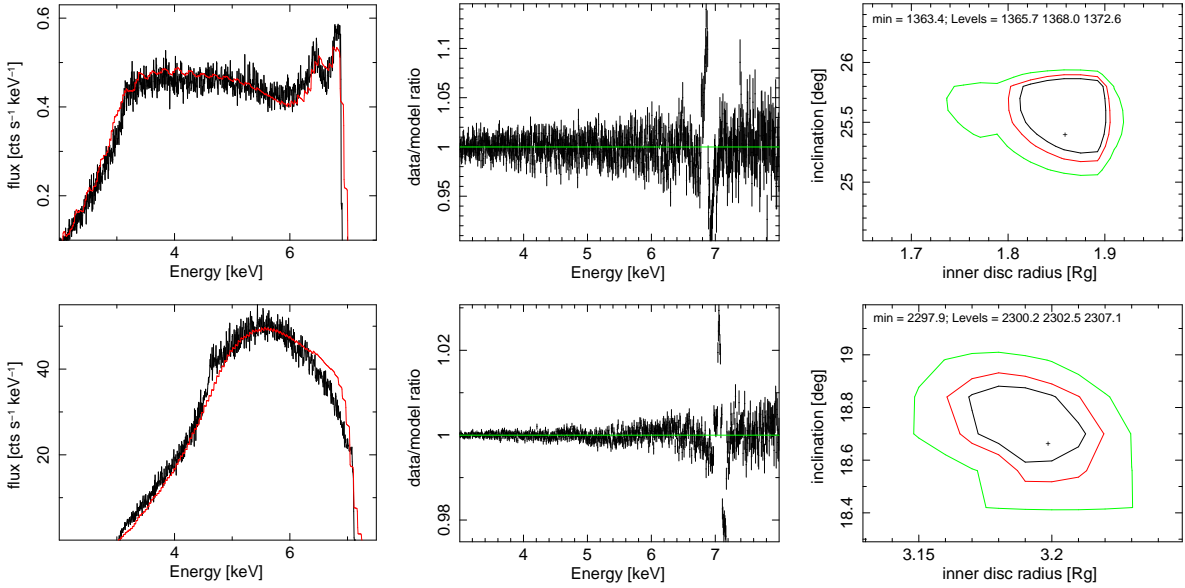

\begin{center}
\begin{tabular}{ccc}
\includegraphics[angle=270,width=0.31\textwidth]{mcg_sim_line.eps} & 
\includegraphics[angle=270,width=0.31\textwidth]{mcg_sim_ratio.eps} &
\includegraphics[angle=270,width=0.31\textwidth]{mcg_sim_contirin_best} \\
\includegraphics[angle=270,width=0.31\textwidth]{gx_sim_line.eps} &
\includegraphics[angle=270,width=0.31\textwidth]{gxf_sim_ratio.eps} &
\includegraphics[angle=270,width=0.31\textwidth]{gx_sim_contirin.eps}\\
\end{tabular}
\caption{The simulated spectra for MCG-6-30-15 (top) and GX 339-4 (bottom) using the preliminary response matrix of the XEUS mission. 
Left: The broad iron line generated
by the {\kyrline} model (black data) and fitted by the {\laor} model (red curve). Middle: The data/model 
ratio when fitting by the {\laor} model. Right: Contours for the inclination angle and the inner disc radius
of the {\laor} model. The trial values are far from the best fit results (see KY value in Table 4).}
\label{xeus_fake}
\end{center}
\end{figure}

Prior to the spectral analysis we rebinned the data to have approximately a 5eV resolution 
(as it was planned for the XEUS instrument). 
We tested different grouping realizing that the discrepancies between the two models increase 
with larger grouping, but we can see apparent differences already for the most moderate rebinning.
The results of the {\laor} fit are shown in Table 4 and in Figure~\ref{xeus_fake}.
The broad iron line component of the model is plotted in the left column. 
The continuum components of the model are not displayed there in order to clearly see the deflections of the {\laor} model.
The most prominent discrepancy appears at the higher-energy drop, which is clearly seen in the data/model ratio plot
in the middle column of Figure~\ref{xeus_fake}. The model parameters are constrained with small error bars 
(see contours $a$ vs. $i$ in the right column of Figure~\ref{xeus_fake}), 
which clearly reveals a difference between the {\kyrline} and the {\laor} models.

The spin value derived from the analysis using the {\laor} model is:
\[
a_{\rm laor,MCG} = 0.958^{+0.003}_{-0.004} 
\]
for MCG-6-30-15, while initially the spin value was $a_{\rm trial,MCG} = 0.940$, and
\[
 a_{\rm laor,GX} = 0.74^{+0.02}_{-0.02} 
\]
for GX 339-4, while initially the spin value was $a_{\rm trial,GX} = 0.70$.\\

\vspace{0.1cm}
\begin{center}
\begin{tabular}{c|c|c||c|c}
 \multicolumn{5}{c}{\bf Table 4. Results of {\laor} fit in 3--9\,keV in the simulated spectra}\\
        \hline \hline
				& \multicolumn{2}{c}{\textbf{MCG-6-30-15}} & \multicolumn{2}{c}{\textbf{GX 339-4}} \\
        parameter               &       KY value        &       fitted value    &       KY value        &       fitted value \\
\hline
	\rule{0cm}{0.35cm}
        $       R_{in} [G]      $&$     2.00     $&$     1.86^{+0.04}_{-0.02}$   &       $3.39$ 	 	&       $3.20^{+0.05}_{-0.05}$   \\ 
        $       i [deg]         $&$     26.7    $&$     25.4(3) $ 		&       $19$    	&       $18.6(2)$      \\
        $       E_{\rm line}        $&$     6.70     $&$     6.71(1) $      	 	&       $6.97$  	&       $6.94(5)$	\\
        $       q_{1}           $&$     4.90     $&$     4.51(3) $       	&       $3.45$  	&       $3.2(1)$        \\
        $       q_{2}           $&$     2.80     $&$     2.76(2) $       	&	-		&	-		\\
        $       r_{b}           $&$     5.5     $&$     6.2(1)  $        	&	-		&	-		\\
        $       K_{\rm line}    $&$     8.7\times 10^{-5}$&$ 9.0\times 10^{-5}$ &       $6.5\times 10^{-3}$&$ 6.6\times 10^{-3}   $\\
\hline
        $\chi^{2}/\nu$            & & $1364/873$    & & $2298/873$             \\
\end{tabular}
\end{center}
\vspace{0.1cm}

\section{Conclusions}
We investigated the iron line band for two representative sources -- MCG-6-30-15 (active galaxy) and GX 339-4 (X-ray binary).
The iron line is statistically better constrained for the active galaxy MCG-6-30-15 due to a significantly longer exposure
time of the available observations -- for comparison of count rates of the sources see Table~3. 
The spectra of both sources are well described by a continuum model 
plus a broad iron line model. We compared modeling of the broad iron line by the two relativistic models, {\laor} and {\kyrline}. 
The {{\kyrline}} model leads to a better defined minimum of $\chi^{2}$ for the best fit value.
The confidence contour plots for $a/M$ versus other model parameters are more regularly shaped.
This indicates that the {{\kyrline}} model has a smoother adjustment between the different
points in the parameter space allowing for more reliable constraints on $a/M$. 
The {{\laor}} model has a less accurate grid and is strictly limited to the extreme Kerr metric. 
The discrepancies between the {{\kyrline}} and {{\laor}} results are within 
the general uncertainties of the spin determination using the skewed line profile
when applied to the current data.
However, the results are apparently distinguishable for higher quality data, as those simulated for the XEUS mission. 
We find that the {\laor} model tends to overestimate the spin value 
and furthermore, it has insufficient energy resolution which affects the correct determination of
the high-energy edge of the broad line.
The discrepancies in the overall shape of the line are more visible especially for lower values of the spin $a/M$.
As a side-product, we have found that the correct re-binning of the data with respect to the instrumental energy resolution
is crucial to obtain statistically the most relevant results.

\acknowledgments 
{The  present  work  was  supported by the ESA Plan for European Cooperating States
(98040).}

\end{article}

\end{document}